\documentclass[11pt,aps,nofootinbib,%twocolumn,
floatfix,superscriptaddress]{revtex4}
\usepackage{graphicx}
\usepackage{epsf,amsmath,amsfonts,amssymb,amsbsy}
\usepackage[mathscr]{eucal}
\usepackage{hyperref}

\begin{document}

\title{\boldmath Mixing of $\eta - \eta'$  in charge-exchange
reactions\\ and decays of mesons with heavy quarks}

\begin{abstract}
{Involving of elastic rescattering and annihilation of quark-antiquark pairs
in a final state allows us to explain a dependence of ratio for cross
sections of $\eta'$ and $\eta$ mesons versus transfer momentum in charge
exchange reactions. We estimate the mixing angle of mesons with isoscalar
states of $\bar u u+\bar d d$ and $\bar s s$ of hidden strangeness. The
evaluation includes the consistent description of yield ratio for $\eta'$ and
$\eta$ mesons in decays of $B^0$, $B^0_s$ and $J/\psi$ mesons.}
\end{abstract}

\author{Ja.V.Balitsky}
\email{BalytskiyYaroslav@gmail.com}
\affiliation{Moscow Institute of Physics
and Technology (State University), Russia, 141701, Moscow Region,
Dolgoprudny, Institutsky 9} \affiliation{Russian State Research Center
Institute for High Energy Physics (National Research Centre Kurchatov
Institute), Russia, 142281, Moscow Region, Protvino, Nauki 1}

\author{V.V.Kiselev}
\email{Valery.Kiselev@ihep.ru} \affiliation{Moscow Institute of Physics and
Technology (State University), Russia, 141701, Moscow Region, Dolgoprudny,
Institutsky 9} \affiliation{Russian State Research Center Institute for High
Energy Physics (National Research Centre Kurchatov Institute), Russia,
142281, Moscow Region, Protvino, Nauki 1}

\author{A.K.Likhoded}
\email{Anatolii.Likhoded@ihep.ru} \affiliation{Moscow Institute of Physics
and Technology (State University), Russia, 141701, Moscow Region,
Dolgoprudny, Institutsky 9} \affiliation{Russian State Research Center
Institute for High Energy Physics (National Research Centre Kurchatov
Institute), Russia, 142281, Moscow Region, Protvino, Nauki 1}

\author{and V.D.Samoylenko}
\email{Vladimir.Samoylenko@ihep.ru}
\affiliation{Russian State Research
Center Institute for High Energy Physics (National Research Centre Kurchatov
Institute), Russia, 142281, Moscow Region, Protvino, Nauki 1}

\maketitle

\section{Introduction}
In the spectroscopy operating by quark quantum numbers, the neutral
pseudoscalar mesons $\eta^{(\prime)}$ can be represented as the
superpositions of isosinglet states
\begin{equation}\label{superposition}
    \begin{array}{l}
    |\eta^\prime\rangle=\sin\phi\,|\bar n n\rangle+\cos\phi\,|\bar s s\rangle,\\[2mm]
    |\eta\,\rangle=\cos\phi\,|\bar n n\rangle-\sin\phi\,|\bar s s\rangle,
    \end{array}
\end{equation}
where  $|\bar s s\rangle$ is composed of strange quark and antiquark, while
$|\bar n n\rangle$ is the isosinglet of light  $u$ and  $d$  quarks:
\begin{equation}\label{isosing}
    |\bar n n\rangle=\frac{1}{\sqrt{2}}\,\left\{|\bar d d\rangle+
    |\bar u u\rangle\right\}.
\end{equation}
In such the representation one suggests the absence of any admixture of
exotic glueball state with the same quantum numbers of parity and isospin
with no valence quarks\footnote{Models, involving the glueball admixing, lead
to suppressed amplitudes of such the admixture
\cite{Novikov:1979uy,Novikov:1979ux,Novikov:1979va,Kataev:1981aw,Kataev:1981gr,
Ambrosino:2009sc,Escribano:2007cd,Escribano:2008rq}. This fact  is natural
because a matrix element of hamiltonian of non-perturbative mixing can be
estimated as $\Lambda_\mathrm{QCD}\sim 300-400$ MeV by the order of
magnitude, while the glueball state takes the mass, which is $2.5-3$ GeV more
heavy than $\eta^{(\prime)}$, at least. Therefore, the amplitude of mixing
with the glueball is about $A_\mathrm{G}\lesssim \Lambda_\mathrm{QCD}/\Delta
M\sim 0.1\ll 1$.}. This simplified representation should be also compared
with a deeper consideration of mixing schemes that use quark currents and
other strict notions, say, in \cite{Kiselev:1992ms,Schechter:1992iz}.

Numerous studies have been devoted to determination of mixing angle $\phi$
for the pseudoscalars, see, for instance, a review of quark model in
\cite{Agashe:2014kda} as well as original articles
\cite{Rosner:1982ey,Bramon:1997va,DiDonato:2011kr,
Feldmann:1998vh,Escribano:2005qq,Huang:2006as,Thomas:2007uy} also including a
study within a holographic approach to quark physics in
\cite{Brunner:2015yha}. In the limit of flavor symmetry for three quarks,
SU$_f$(3), one can evidently get $\sin\phi\mapsto 2/\sqrt{6}$, so that
$\eta'\to\eta_0$, $\eta\to \eta_8$, i.e. one arrives to SU$_f$(3) singlet and
SU$_f$(3) octet, respectively:
$$
    \begin{array}{ll}
    |\eta_0\rangle=\frac{1}{\sqrt{3}}&\left\{|\bar u u\rangle+
    |\bar d d\rangle+\,|\bar s s\rangle\right\},\\[2mm]
    |\eta_8\,\rangle=\frac{1}{\sqrt{6}}&\left\{|\bar u u\rangle+
    |\bar d d\rangle-2\,|\bar s s\rangle\right\}.
    \end{array}
$$
Such the rotation of neutral states of SU$_f$(3) singlet  and  SU$_f$(3)
octet to the isosinglet states $|\bar n n\rangle$ and $|\bar s s\rangle$ with
the angle of $\cos^2\phi_\star=\frac13$ is usually called the ideal mixing:
\mbox{$\phi_\star=\arctan\sqrt{2}\approx 54.7^o$.} One usually introduces the
mixing angle $\theta$ by $\theta=\phi-\phi_\star\approx\phi-54.7^o$.

The difficulty of problem on the mixing is caused by the issue on how the
mixing introduced in the spectroscopy is implemented in the description of
dynamical processes and can been measured in a model-independent way. For
instance, the measurement of cross section ratio for $\eta^{(\prime)}$ mesons
in the charge exchange reactions gives the result dependent on the transfer
momentum $t$ \cite{Donskov:2013uva}. It is problematic to make this fact
straightforwardly compatible with simple spectroscopic notions, since the
spectroscopy operates by some constant quantities, but dynamical functions.

In the present paper we show that the dynamical characteristics in the charge
exchange reactions with the production of $\eta^{(\prime)}$ are really
consistent with the picture of meson mixing if we introduce the interaction
in the final states for the isosinglets with taking into account the
annihilation of quark-antiquark pairs. A relative contribution of such the
annihilation acquires reasonable features in its magnitude and behavior with
respect to the momentum transfer. This allows us to constrain a data region,
wherein the mixing angle can be measured in the model-independent way (see
Section \ref{AmAn}).

Further, in Sections \ref {Psi} and \ref{B} we show that the measurements of
$\eta^{(\prime)}$ yields in radiative decays of $J/\psi$ as well as in some
channels of decays of neutral $B$ mesons are also consistent with the
mechanism taking into account the rescattering of quark-antiquark pair in the
final state.
\section{Amplitude analysis of  $t$-dependence for ratio of yields of
$\eta$ and $\eta'$ mesons in charge exchange reactions of $\pi^-p$ and $K^-p$
\label{AmAn} }

\begin{figure}[ht]
\setlength{\unitlength}{1mm}
\begin{center}
\begin{picture}(110,27)
\put(0,0){  \includegraphics[width=100\unitlength]{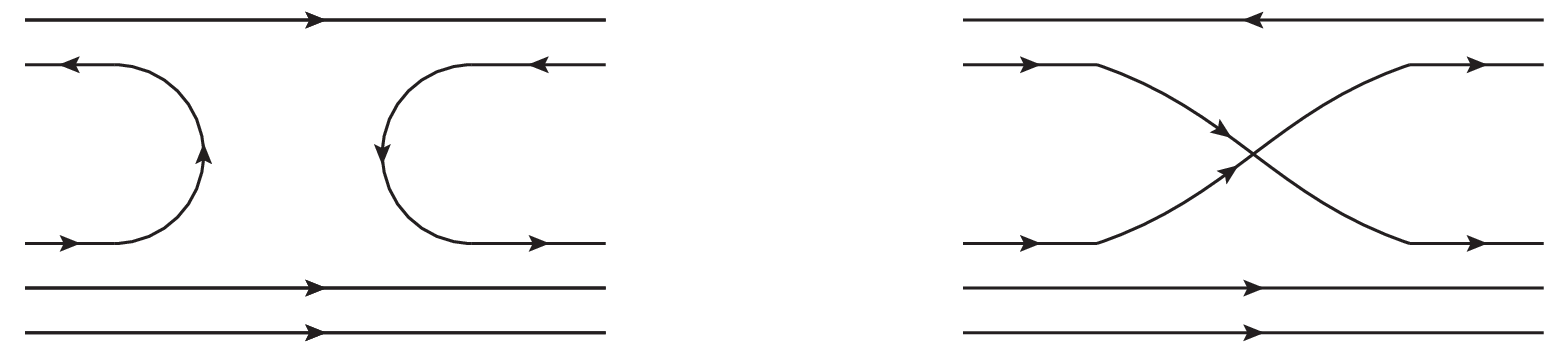}}
\put(20,23){$d$}
\put(10,12){$u$}
\put(28,12){$d$}
\put(20,6){$d$}
\put(20,-2){$u$}
%%%%
\put(80,23){$u$}
\put(70,9){$u$}
\put(88,9){$d$}
\put(80,6){$d$}
\put(80,-2){$u$}
\end{picture}
\end{center}
  \caption{The exchange by quark flavors in the process of $\pi^-p\to n\eta(\eta')$
with the production of $|d\bar d\rangle$ and $|u\bar u\rangle$ states.}\label{f1}
\end{figure}

The exchange by quantum numbers of flavors in the reactions $\pi^-p\to n\eta
(\eta')$ is shown\footnote{See \cite{Nekrasov:2013dba} about taking into
account both terms in Fig.\ref{f1}. } in Fig.~\ref{f1}. It corresponds to
Regge's exchange with the trajectory possessing the quantum numbers of $a$
mesons. The amplitude of $\mathcal M$ corresponds to the contribution when
$d$ quark is a spectator, while $\mathcal M_B$ denotes the term when the
spectator is $\bar u$ antiquark.

\begin{figure}[ht]
\setlength{\unitlength}{1.2mm}
\begin{center}
\begin{picture}(110,37)
\put(0,0){  \includegraphics[width=100\unitlength]{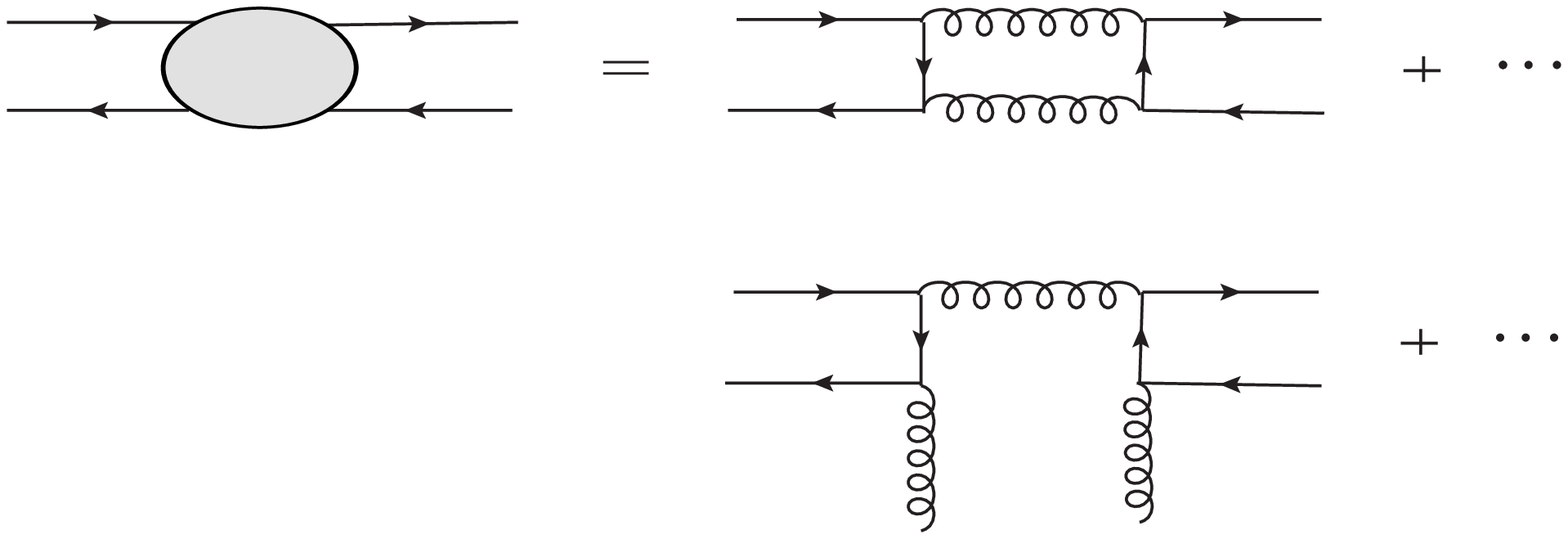}}
\put(7,35){$q$}
\put(7,24){$q$}
\put(27,35){$q'$}
\put(27,24){$q'$}
%%%%%%%%%%
\put(53,35){$q$}
\put(53,24){$q$}
\put(79,35){$q'$}
\put(79,24){$q'$}
%%%%%%%%%%
\put(53,18){$q$}
\put(53,7){$q$}
\put(79,18){$q'$}
\put(79,7){$q'$}
\end{picture}
\end{center}
\caption{The interaction in the final state of $q\bar q$ pair with the
annihilation.}
\label{f2}
\end{figure}

The quark-antiquark state produced in the charge exchange reaction has the
invariant mass in the range of $500-1000$ MeV, hence, there is a strong
non-perturbative interaction in the final state. In the case of pseudoscalar
color-singlet channel the interaction permits the annihilation of
quark-antiquark state into the quark-antiquark state, say, by the pair of
gluons or the one-gluon annihilation under the two-gluon exchange with the
baryon (see Fig.~\ref{f2}), if the quark flavor is changed. Analogously,
there should be the elastic scattering if the quark flavors are conserved. In
this way, the two-gluon exchange with the baryon that corresponds to the
pomeron exchange, leads to a $t$-dependence of the amplitude of interaction
in the final state. Therefore, taking into account the interaction in the
final state as pictured in Fig.~\ref{f2} corresponds to the contribution of
double reggeon exchange, namely, the reggeon-pomeron term.

The charge-exchange amplitude of $\pi^-p\to\eta(\eta')n$ taking into account
the interaction in the final state of $q \bar q$ pair with the possibility of
annihilation (marked by subscript $\mathscr A$) as well as the rescattering
takes the general form
\begin{equation}{\label{revis1}}%$$
\begin{array}{rl}
    \mathcal M[\pi]=&%\left\{
    \mathscr M |d\bar d\rangle+
    \mathscr M_A (|d\bar d\rangle+|u\bar u\rangle+
    \sqrt{\lambda_s}|s\bar s\rangle)%\right\}
    + %\left\{
    \mathscr M_B |u\bar u\rangle+
    \mathscr M_{AB} (|d\bar d\rangle+|u\bar u\rangle+
    \sqrt{\lambda_s}|s\bar s\rangle)%\right\}
    \\[3mm]
    =&               \mathscr M
    \left\{|d\bar d\rangle+
    \mathscr A (|d\bar d\rangle+|u\bar u\rangle+
    \sqrt{\lambda_s}|s\bar s\rangle)
    +\mathscr B\left\{ |u\bar u\rangle+
    \mathscr A (|d\bar d\rangle+|u\bar u\rangle+
    \sqrt{\lambda_s}|s\bar s\rangle)\right\}\right\},
\end{array}
\end{equation}%$$
where  $\lambda_s$ is the parameter of suppression for the strange quark
production at the given invariant masses.

In this representation of amplitude we ignore the polarization effects
related to spins of nucleons and an orientation of scattering plane. These
effects can compose several percents of differential cross section, but they
are not essential for our purpose to isolate main regularities connected to
the mixing and t-dependence of measured  quantities. Anyway, the offered
amplitude is inevitably dictated by the processes with the quark quantum
numbers, i.e. by the quark scheme of hadrons itself. In this respect, we
label the quark amplitudes by the valence quark quantum states, $|q
q'\rangle$, for the visual clarity of notions.

For the sake of simplicity, we put $\lambda_s$ to be real, constant and equal
to $\lambda_s=\frac12$. These properties of parameter $\lambda_s$ can be
justified in the following way:

First, the parameter describes the relative fraction of strange
quark-antiquark pair creation in the fixed range of invariant masses
corresponding to the mass interval of neutral mesons, so that we could
suggest its value to be a constant.

Second, the strange quark fraction in the annihilation channel is independent
on the transverse momentum, since such the dependence is factorized, and it
can be suggested to be identical for different flavors of light quarks in the
region of invariant masses greater than the current masses of light quarks,
while the only difference for the strange quarks is given by the suppression
of overall probability for the creation of strange quark-antiquark pair in
the mentioned interval of invariant masses.

Third, the magnitude of $\lambda_s$ can be evaluated empirically in other
processes with the same mechanism with the annihilation channel in the final
state interaction for the quark amplitudes. Such the extraction of
$\lambda_s$ is presented in next Section, wherein $\eta^{(\prime)}$ are
produced in the radiative decays of $J/\psi$. We will see that the
uncertainty of $\lambda_s$ is numerically significant, but, it is important,
its variation certainly belongs to the range less than unit and the
introduction of $\lambda_s$ is principal with no doubt. So, the chosen value
of $\lambda_s$ is consistent with the other processes, while its variation
due to the uncertainty will change the model parameters, of course, but this
variation does not change our basic conclusions as we will see.

The ratio of $\eta$ and $\eta'$ yields does not depend on factor $\mathscr
M$, so that the overall behavior of this ratio is determined by complex
amplitudes $\mathscr A$ and $\mathscr B$. In this way, in the case of $\pi^-
p$ process the result does not depend on $\mathscr B$, and the $t$-dependence
of the ratio is completely given by the function of $\mathscr A(t)$.

Indeed, the expansion in terms of the isosinglet state $|\bar n n\rangle$
according to (\ref{isosing}) and the state with hidden strangeness $|s\bar
s\rangle$ gives
\begin{equation}{\label{revis2}}%$$
\mathcal M[\pi]\sim (1+\mathscr
B)\left\{ \frac{1}{\sqrt{2}}\,(1+2\mathscr A)\,|n\bar n\rangle+ \mathscr
A\,\sqrt{\lambda_s}\,|s\bar s\rangle \right\},
\end{equation}%$$
hence,
\begin{equation}{\label{revis3}}%$$
    R_\pi=\left|\frac{\mathcal M[\pi\to\eta']}
    {\mathcal M[\pi\to\eta]}\right|^2=
    \left|\frac{\frac{1}{\sqrt{2}}\,(1+2\mathscr A)\,\tan \phi+
    \mathscr A\,\sqrt{\lambda_s}}{
    \frac{1}{\sqrt{2}}\,(1+2\mathscr A)-
    \mathscr A\,\sqrt{\lambda_s}\,\tan \phi}
    \right|^2.
\end{equation}%$$
The exclusion of the final state interaction in the quark-antiquark state
with the annihilation, i.e. $\mathscr A\equiv 0$, leads to the quantity
independent of $t$:
$$
    R_\pi\to \tan ^2\phi,
$$
that contradicts to the experimental data. This fact directly shows that the
amplitude of such interaction in the final state is not equal to zero.
Moreover, at fixed mixing angle we can restore a form of function $\mathscr
A(t)$ by assuming a model for its complex phase.

\begin{figure}[ht]
\setlength{\unitlength}{1mm}
\begin{center}
\begin{picture}(110,27)
\put(0,0){  \includegraphics[width=100\unitlength]{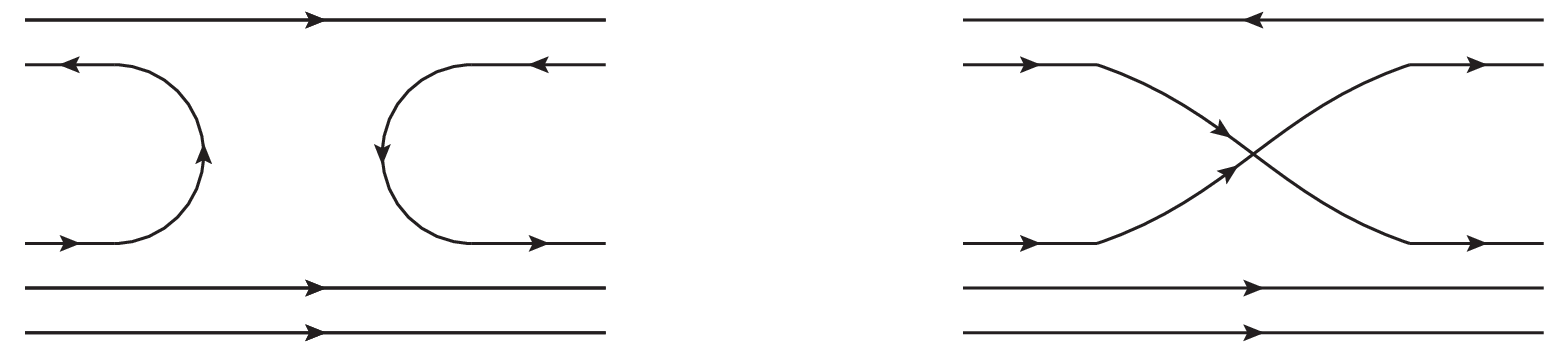}}
\put(20,23){$s$}
\put(10,12){$u$}
\put(28,12){$s$}
\put(20,6){$d$}
\put(20,-2){$u$}
%%%%
\put(80,23){$u$}
\put(70,9){$u$}
\put(88,9){$s$}
\put(80,6){$d$}
\put(80,-2){$u$}
\end{picture}
\end{center}
  \caption{The exchange by quark flavors in the process of
  $K^-p\to \Lambda\eta(\eta')$ with the production of
  $|s\bar s\rangle$ and $|u\bar u\rangle$ states.
}
  \label{f3}
\end{figure}

Analogously, in accordance with diagrams in Fig.~\ref{f3} the amplitude of
charge exchange reaction $K^-p\to \Lambda \eta(\eta')$ takes the
form\footnote{The term $\mathcal M'$ corresponds to the contribution when the
$s$ quark is the spectator, while $\mathcal M_B'$ corresponds to the term
with the spectator $\bar u$. The amplitude of annihilation channel is marked
by subscript $A$.}
\begin{equation}{\label{revis4}}%$$
\begin{array}{rl}
    \mathcal M[K]=&%\left\{
    \mathscr M' |s\bar s\rangle+
    \mathscr M_A' (|d\bar d\rangle+|u\bar u\rangle+
    \sqrt{\lambda_s}|s\bar s\rangle)%\right\}
    +%\left\{
    \mathscr M_B' |u\bar u\rangle+
    \mathscr M_{AB}' (|d\bar d\rangle+|u\bar u\rangle+
    \sqrt{\lambda_s}|s\bar s\rangle)%\right\}=
    \\[3mm]
    =&               \mathscr M'
    \left\{|s\bar s\rangle+
    \mathscr A (|d\bar d\rangle+|u\bar u\rangle+
    \sqrt{\lambda_s}|s\bar s\rangle)
    +\mathscr B\left\{ |u\bar u\rangle+
    \mathscr A (|d\bar d\rangle+|u\bar u\rangle+
    \sqrt{\lambda_s}|s\bar s\rangle)\right\}\right\},
\end{array}
\end{equation}%$$
wherein we suggest the invariance of high energy amplitudes with respect to the
transformation of light quark flavors.
Then,
\begin{equation}{\label{revis5}}%$$
\mathcal M[K]\sim
\frac{1}{\sqrt{2}}\,\left\{\mathscr B+2\mathscr A\,(1+\mathscr B)\right\}
\,|n\bar n\rangle+
\left\{1+\mathscr A\,\sqrt{\lambda_s}(1+\mathscr B)\right\}\,|s\bar s\rangle,
\end{equation}%$$
hence,
\begin{equation}{\label{revis6}}%$$
    R_K=\left|\frac{\mathcal M[K\to\eta']}
    {\mathcal M[K\to\eta]}\right|^2=
    \left|\frac{\frac{1}{\sqrt{2}}\,
    \left\{\mathscr B+2\mathscr A\,(1+\mathscr B)\right\}\,\tan \phi+
    \left\{1+\mathscr A\,\sqrt{\lambda_s}(1+\mathscr B)\right\}}{
    \frac{1}{\sqrt{2}}\,\left\{\mathscr B+2\mathscr A\,(1+\mathscr B)\right\}-
    \left\{1+\mathscr A\,\sqrt{\lambda_s}(1+\mathscr B)\right\}\,\tan \phi}
    \right|^2.
\end{equation}%$$
We assume that the absolute value of $\mathscr B$  weakly depends on $t$ at
high energies, so that in calculations we put $\mathscr B$ equal to a complex
number with a phase which can depend on $t$. Such the prescription allows us
to study the dependence of amplitude on a single complex function $\mathscr
A(t)$. This assumption is quite rigid and it restricts a possibility to fit
the experimental data, because the extraction of function $\mathscr A(t)$
from $R_\pi$ fixes a value of $R_K$. However, the assumption is reasonable,
since it allows us quite satisfactory to describe the behavior of $R_K$.

Notice that in eqs. (\ref{revis4})--(\ref{revis6}) for the $K^-p$ scattering
we have used the same notation $\mathscr B$ for the fraction of amplitude
with the spectator $u$-quark as it has been involved in the description of
$\pi^- p$ scattering, since for the sake of brevity we have missed the prime
for this quantity in eqs. (\ref{revis4})--(\ref{revis6}), because the ratio
of cross sections under study in $\pi^- p$ charge exchange reaction does not
depend on $\mathscr B$ at all\footnote{Therefore, we have not deal with an
assumption about the SU(3)-flavor symmetry for the fraction $\mathscr B$, it
is just the economy of notations.}.

Moreover, we remind that in the Regge calculus the two processes with the
different spectator quarks corresponds to the identical Regge trajectories,
hence, we deal with the quark amplitude decomposition of the same amplitude.
Therefore, it would be natural to expect that two quark amplitudes have got
the same $t$-dependence, while their ratio in terms of factor $\mathscr B$
does not depend on the transverse momentum at all, since we consider the same
decomposition of unified Regge amplitude at different $t$ in terms of two
amplitudes with the definite exchange of quark quantum numbers. Thus, our
assumption on the negligible $t$-dependence of fraction $\mathscr B$ is
justified and natural.

Let us note in the very beginning of analysis that we consider two versions of
$\mathscr A(t)$ behavior: $\mathscr A(t)$ essentially grows or falls with
the momentum transfer increase.

\subsection{\textit {Scenario I}}

The significant increasing of $\mathscr A(t)$ means that this contribution is
practically negligible at low momentum transfers $t\to t_\mathrm{min}$, where
$t_\mathrm{min}\to 0$ at high energies of scattering, i.e. at energies much
greater than masses of particles in the process, $s\gg m^2$, so that in what
follows we operate with a formal limit of $t\to 0$. In this way, the
contribution of $\mathscr A(t)$ becomes significant only at momentum
transfers comparable with the particle masses squared. Under such the
assumptions the tangent of mixing angle is determined by the factor of
$R_\pi=\tan ^2\phi$ at zero momentum transfer, and it is quite large, $\tan
\phi\approx 0.75$. This version of behavior should lead to both a fall of
$R_\pi$ factor and a reasonable value of $R_K$ raising versus $|t|$. We find
that this scenario can be actualized if the following two conditions take
place: at first, the imaginary part of factor $\mathscr A(t)$ is negligibly
small and, second, the absolute value and phase of $\mathscr B$ at $t\to 0$
are adjusted in order to describe $R_K(t\to 0)$, while at $t\neq 0$ the phase
of $\mathscr B$ depends on the momentum transfer in the way to fit the $R_K$
data at constant fixed absolute value of $\mathscr B$.

\begin{figure}[th]
\setlength{\unitlength}{0.9mm}
\begin{center}
\begin{picture}(160,55)
\put(0,0){  \includegraphics[width=75\unitlength]{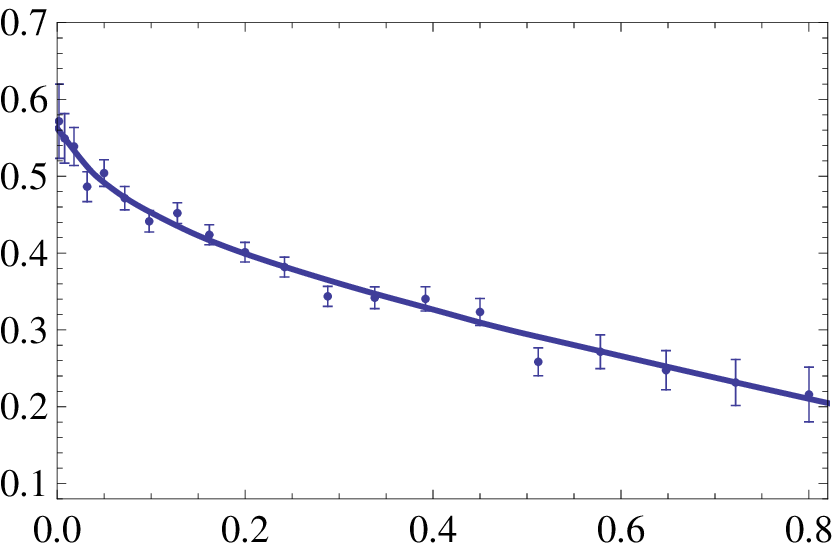}}
\put(80,0){  \includegraphics[width=75\unitlength]{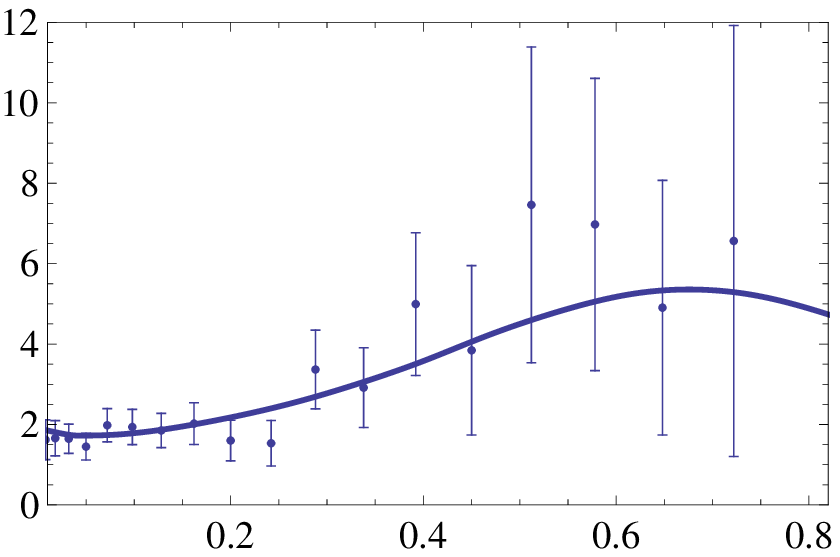}}
\put(60,-4){$-t$, GeV$^2$}
\put(140,-4){$-t$, GeV$^2$}
\put(0,52){$R_\pi$}
\put(80,52){$R_K$}
\end{picture}
\end{center}
  \caption{The description of charge exchange data \cite{Donskov:2013uva}
  in Scenario I.}
  \label{mI}
\end{figure}

For instance, we find the model I
\begin{equation}{\label{revis7}}%$$
    \tan \phi=0.75,\quad\mathscr  B=\frac89\,\mathrm{e}^{-i\pi/2.3
    (1-f(a))},
    \quad \mathscr  A=-a,\quad f(a)=\frac{5(5a)^{2.85}}{1+3(5a)^{2.85}},
\end{equation}%$$
where $a$ is a real positive parameter corresponding to the relative
contribution of annihilation channel. The quality of data description in this
model\footnote{We do not show a statistical likelihood of the model because
we have aimed to demonstrate the principal importance of annihilation channel
in the final state. In addition, a justification of amplitude values
extracted in the non-perturbative regime seems to be very problematic, so
that the procedure of fitting the data could be improved to the ideal
agreement without achieving a distinguished result.} is shown in
Fig.~\ref{mI}.

\begin{figure}[th]
\setlength{\unitlength}{0.9mm}
\begin{center}
\begin{picture}(100,55)
\put(0,0){  \includegraphics[width=75\unitlength]{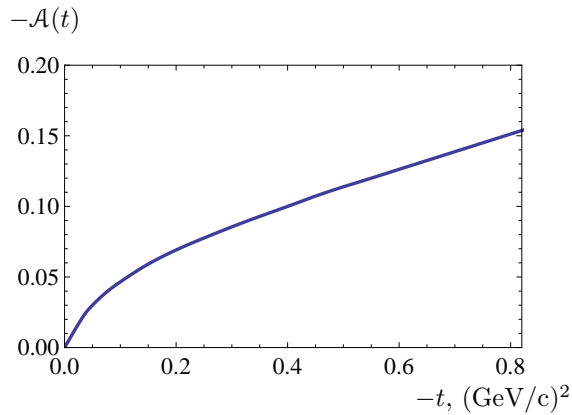}}
\put(60,-4){$-t$, (GeV/c)$^2$}
\put(0,52){$-\mathscr A(t)$}
\end{picture}
\end{center}
  \caption{The dependence of $\mathscr A(t)$ versus the momentum transfer in
 model I.}
  \label{mI-a}
\end{figure}

The behavior of  $\mathscr A(t)$ (or simply the connection of parameter $a$
to the transfer momentum) in Scenario I is shown in Fig.~\ref{mI-a}. The
characteristic values of amplitude $\mathscr A(t)$ determine the magnitude of
violation of OZI rule: the production of valence quark-antiquark pairs is
suppressed with respect to the processes without such the production. The
accuracy of such approximation is about 10\% in the amplitude, hence, about
1\% in the probability, if the interference with the leading term is
suppressed by relevant quantum numbers.

\subsection{\textit{Scenario II}}

In this case, the factor of $\mathscr A(t)$ significantly decreases with the
growth of momentum transfer $|t|$. In this way, $\tan ^2\phi$ represents the
bottom limit for $R_\pi$. i.e. it gets small values saturating $R_\pi$ at
high momentum transfer. Currently, the degree of such the saturation can not
be strictly established from the data because of valuable experimental
uncertainties at high momentum transfers. Nevertheless, we can state that the
small mixing angle in Scenario II would fit the experimental data with a low
confidence level.

\begin{figure}[th]
\setlength{\unitlength}{0.9mm}
\begin{center}
\begin{picture}(160,55)
\put(0,0){  \includegraphics[width=75\unitlength]{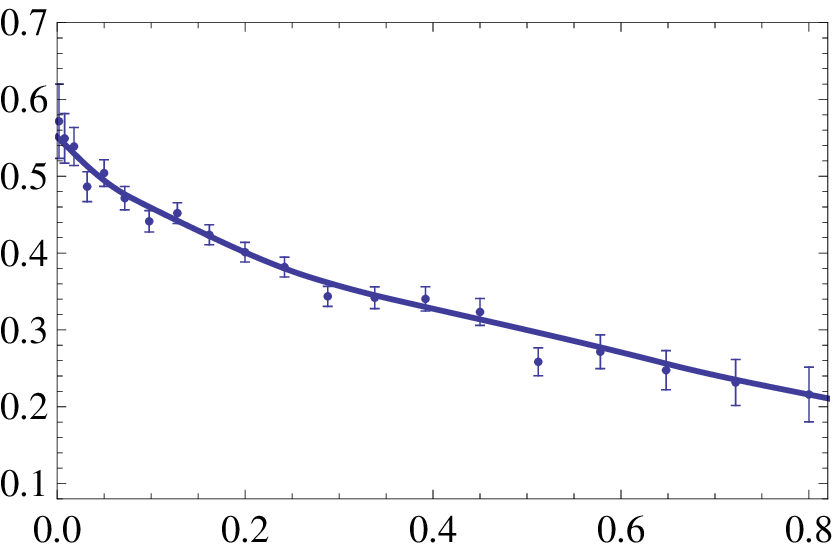}}
\put(80,0){  \includegraphics[width=75\unitlength]{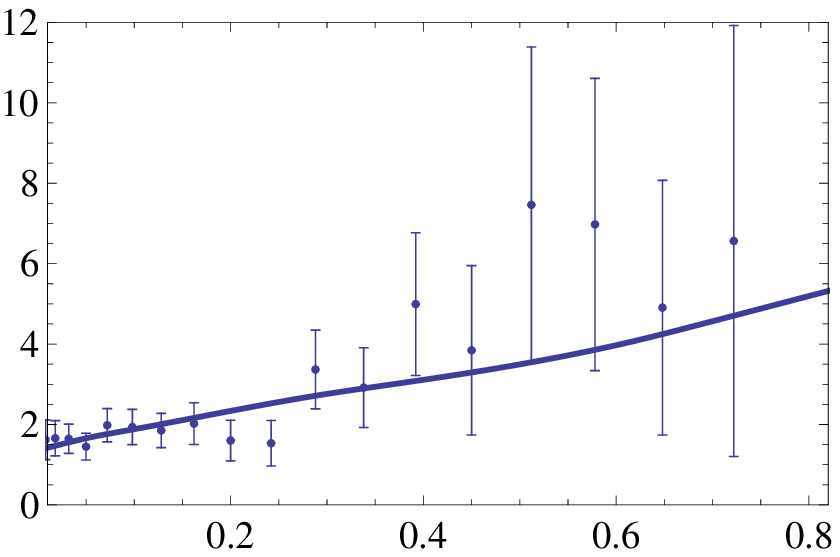}}
\put(60,-4){$-t$, GeV$^2$}
\put(140,-4){$-t$, GeV$^2$}
\put(0,52){$R_\pi$}
\put(80,52){$R_K$}
\end{picture}
\end{center}
  \caption{The description of charge exchange data \cite{Donskov:2013uva}
  in Scenario II.}
  \label{mII}
\end{figure}

The most simple modeling corresponds to a constant phase of amplitude
$\mathscr A(t)$ equal or close to  $\frac12 \pi$, so that $R_\pi$  decreases
with the momentum transfer growth. The opportunity of acceptable data
description for $R_K$ at $\mathscr  B\equiv 0$ can be gotten in model II (see
Fig. \ref{mII}):
\begin{equation}{\label{revis8}}%$$
    \tan \phi=0.245,\quad\mathscr  B\equiv 0,
    \quad \mathscr  A=-a\,\mathrm{e}^{i\pi/2},\quad 0.33< a < 0.97.
\end{equation}%$$

\begin{figure}[t]
\setlength{\unitlength}{0.9mm}
\begin{center}
\begin{picture}(100,55)
\put(0,0){  \includegraphics[width=75\unitlength]{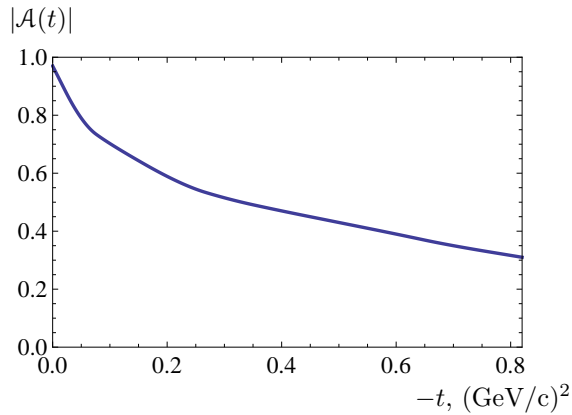}}
\put(60,-4){$-t$, (GeV/c)$^2$}
\put(0,52){$|\mathscr A(t)|$}
\end{picture}
\end{center}
  \caption{The dependence of $\mathscr A(t)$ versus the momentum transfer in
 model II.}
  \label{mII-a}
\end{figure}

Since the region of data is constarined by $|t|< 0.8$ GeV$^2$, model II does
not reach the limit case of $\mathcal A\to 0$ at large values of $|t|$,
corresponding to the saturation of ratio $R_\pi\to \tan^2\phi$. Moreover, in
this scenario the value of $|\mathscr A|$  corresponds to the magnitude of
OZI rule breaking, and it takes an extremely unreal value of the order of
unit at small transverse momenta (see Fig. \ref{mII-a}).

It is worth to note that a variation of parameters in scenario II  does not
allows us to get the tangent of mixing angle greater than $\frac12$.
\subsection{Inference}

The physical contents of Scenarios described above differ essentially: in
version I, the elastic rescattering dominates over the annihilation
interaction at $t\to 0$, so that the ratio of amplitudes, i.e. $\mathscr
A(t)$ is suppressed and it grows versus the momentum transfer, while in
version II the role of elastic channel increases with the momentum transfer
$|t|$. To our opinion, Scenario I is the most realistic. Therefore, the ratio
of cross sections for $\eta^{(\prime)}$ in the charge exchange reaction of
$\pi p $  at $t\to 0$ gives the measured angle of mixing. The description of
mixing in the $K$ reaction is model dependent.

Note that the main result of our study is the decomposition of charge
exchange amplitude versus the quark quantum numbers with the involvement of
annihilation fraction in the final state interaction, that cannot be
neglected. We do empirically extract these decompositions in the case of
kinematics of charge exchange reactions. The decomposition itself is
universal. However, the same approach if used in other processes with the
production of $\eta^{(\prime)}$-mesons with a different kinematics should
result in a change of decomposition amplitudes, of course, as we will see in
next Sections devoted to decays of heavy mesons, when one cannot talk on the
t-channel of amplitude at all.

\section{\boldmath Decays of $J/\psi\to\gamma\eta^{(\prime)}$ \label{Psi}}

\begin{figure}[th]
\setlength{\unitlength}{0.7mm}
\begin{center}
\begin{picture}(100,55)
\put(0,0){  \includegraphics[width=75\unitlength]{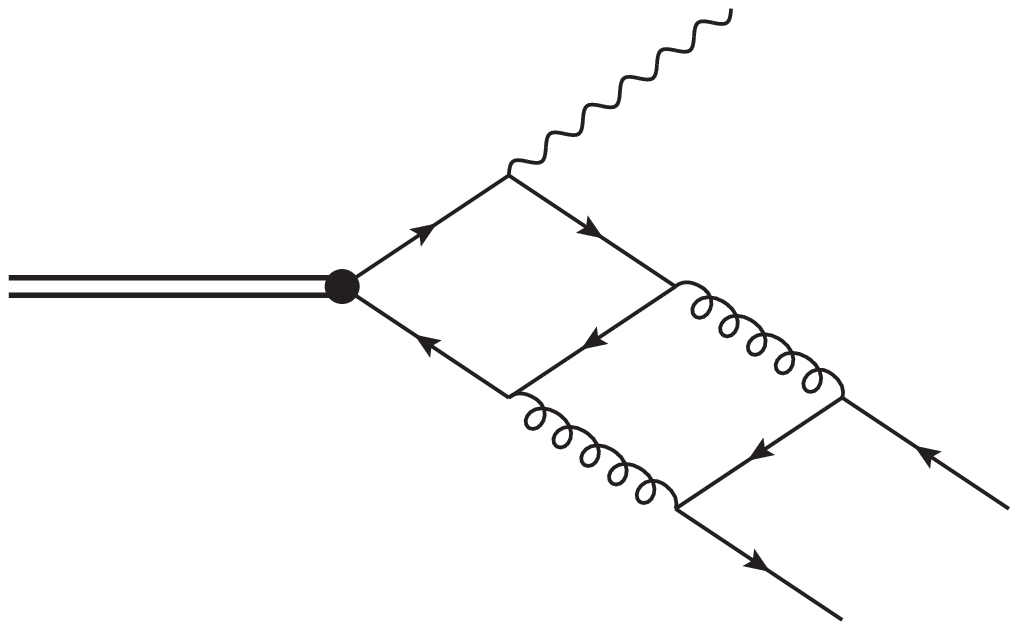}}
\put(55,45){$\gamma$}
\put(5,29){$J/\psi$}
\put(30,30){$c$}
\put(30,17){$\bar c$}
\put(64,-1){$u\,(d,\,s)$}
\put(76,7){$\bar u\, (\bar d,\,\bar s)$}
\end{picture}
\begin{picture}(100,55)
\put(0,15){  \includegraphics[width=85\unitlength]{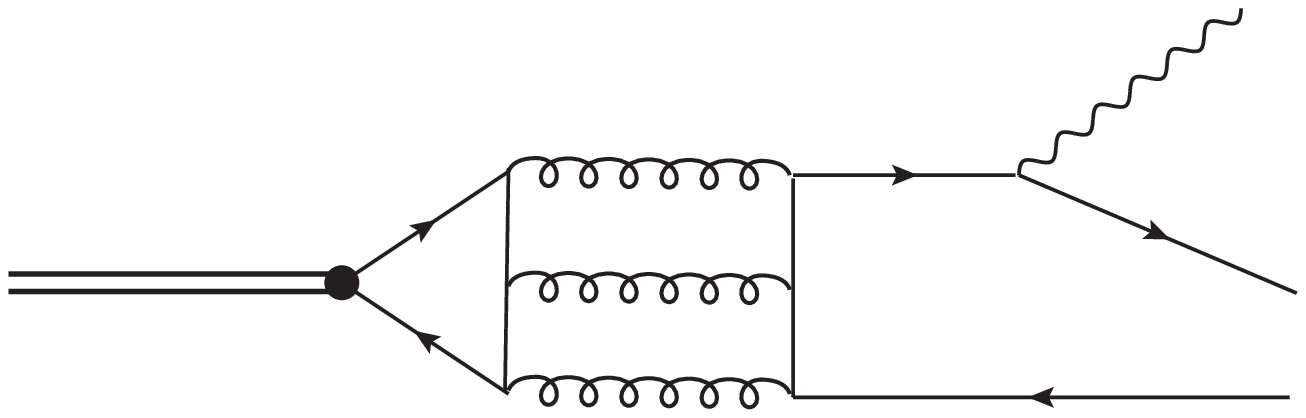}}
\put(83,43){$\gamma$}
\put(5,29){$J/\psi$}
\put(27,29){$c$}
\put(28,15){$\bar c$}
\put(82,25){$u\,(d,\,s)$}
\put(82,12){$\bar u\, (\bar d,\,\bar s)$}
\end{picture}
\end{center}
  \caption{Two kinds of diagrams for the radiative decays
$J/\psi\to\gamma\eta^{(\prime)}$ with the annihilation of two charmed quarks
into the light quarks. }
  \label{rad-psi}
\end{figure}

In radiative decays of $J/\psi$ to $\eta^{(\prime)}$, the mesons are produced
via the annihilation, only (see Fig.~\ref{rad-psi}). Two different kinds of
diagrams are in action: the first is the radiative transition of $c\bar c$
pair to the pseudoscalar state with further annihilation via the channel with
the quantum numbers of two gluons (the left diagram in Fig.~\ref{rad-psi}),
and the second is the mixing of vector $c \bar c$ state with the vector state
composed of light quarks in the channel with quantum numbers of three gluons
with the further radiative transition of vector state into the pseudoscalar
meson (the right diagram in Fig.~\ref{rad-psi}). The second mechanism
involves the breaking down the isospin symmetry by the electromagnetic
interaction, since the electric charges of $u$ and $d$ quarks are different.
We suppose that this contribution to the isospin-symmetry breaking is
irrelevant to our consideration based on the dominance of exact isospin
symmetry, though more careful study was performed in \cite{Gerard:2013gya},
wherein both kinds of diagrams are discussed in the the respect of mixing
problem\footnote{The paper \cite{Gerard:2013gya} has been also focused on the
radiative $\psi'$ decays.}. Therefore, we neglect the mixing of $J/\psi$ with
the light vector states\footnote{Otherwise, the isospin-symmetry breaking
effects would be of the order of unit in the decays under consideration. In
\cite{Gerard:2013gya} the suppression of magnitude for the contribution of
isospin-symmetry breaking in the radiative $j/\psi$ decays is estimated by
the factor greater than 10.} and make calculations for the dominant diagram
of first kind.

%\noindent
In the mechanism taking into account the suppression of the strange
quarks, that violates the flavor symmetry of SU$_f(3)$, one can easily find
that the ratio of $\eta^{(\prime)}$ yields gets the form\footnote{See
\cite{Bramon:2000fr}, wherein analogous radiative decays of light vector
mesons are also considered.}
\begin{equation}\label{rpsi}
    R_\psi=\frac{\Gamma[J/\psi\to\gamma\eta']}{\Gamma[J/\psi\to\gamma\eta]}=q
    \left|\frac{\sqrt{2}\tan\phi+\sqrt{\lambda_s}}
    {\sqrt{2}-\sqrt{\lambda_s}\tan\phi}\right|^2,
\end{equation}
wherein we define the factor
$$
    q=\left(\frac{k_\gamma[J/\psi\to\eta^\prime]}
    {k_\gamma[J/\psi\to\eta]}\right)^3,
$$
which is caused by differences in phase spaces and matrix elements squared
and it is proportional to the third degree of photon momentum $k_\gamma$ in
the rest frame of $J/\psi$. Numarically, $q\approx 0.81$, that indicates
significant correction. Then, the ratio $R_\psi$ essentially depends on the
parameter of strange quark suppression, $\lambda_s$, while the accuracy of
experimental value $R_\psi$ is much less than the accuracy of empirical
estimate for $\lambda_s$.
\begin{figure}[h]
\setlength{\unitlength}{1.1mm}
\begin{center}
\begin{picture}(100,51)
\put(0,0){  \includegraphics[width=75\unitlength]{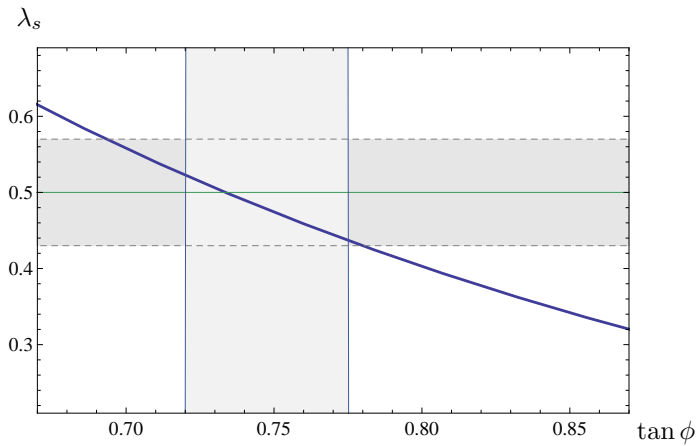}}
\put(2,50){$\lambda_s$}
\put(77,0){$\tan\phi$}
\end{picture}
\end{center}
  \caption{The dependence of suppression factor $\lambda_s$ for the
  strange sea versus the mixing angle $\phi$  according to (\ref{rpsi})
  at the fixed experimental value of $R_\psi$. The vertical shaded band
  restricts the region of tangent for the mixing angle as measured in the
  charge exchange reaction.}
  \label{psi-rate}
\end{figure}

Therefore, it is worth to draw the function  connecting $\tan \phi$ with
$\lambda_s$ according to (\ref{rpsi}). We show this function in
Fig.~\ref{psi-rate}, wherein we have pictured the region of variation for
$\lambda_s=0.5\pm0.07$ as one has usually adopted in the phenomenology
involving the usage of quantity $\lambda_s$.

Thus, we get
\begin{equation}\label{tg-psi}
    \tan\phi=0.733\pm0.045,
\end{equation}
which is in agreement with the value of mixing angle found in the charge
exchange reactions above. However, the overlap with the region extracted from
the charge exchange reactions results in the more strict estimate
\begin{equation}\label{strict}
    \tan\phi=0.740\pm0.022.
\end{equation}

\section{\boldmath Decays of $B^0_{(s)}\to J/\psi\,\eta^{(\prime)}$ \label{B} }

\begin{figure}[ht]
\setlength{\unitlength}{0.7mm}
\begin{center}
\begin{picture}(100,55)
\put(0,0){  \includegraphics[width=75\unitlength]{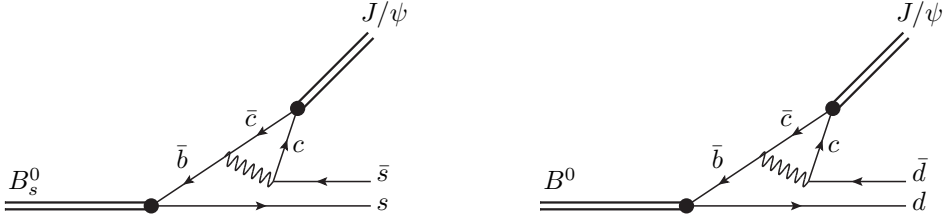}}
\put(72,38){$J/\psi$}
\put(5,6){$B_s^0$}
\put(37,10){$\bar b$}
\put(75,2){$s$}
\put(75,7.5){$\bar s$}
\put(59,13){$c$}
\put(50,18){$\bar c$}
%\put(47,10){$W$}
\end{picture}
\begin{picture}(100,55)
\put(0,0){  \includegraphics[width=75\unitlength]{B-psi}}
\put(72,38){$J/\psi$}
\put(5,6){$B^0$}
\put(37,10){$\bar b$}
\put(75,2){$d$}
\put(75,7.5){$\bar d$}
\put(59,13){$c$}
\put(50,18){$\bar c$}
\end{picture}
\end{center}
  \caption{Diagrams of weak decays $B^0_{(s)}\to J/\psi\,\eta^{(\prime)}$,
  wherein one has to take into account the contribution of
  further annihilation channel for $s\bar s$ and $d\bar d$ pairs.}
  \label{B-psi}
\end{figure}

In weak decays of neutral $B$ mesons transformed to the charmonium $J/\psi$
in association with $\eta^{(\prime)}$, different pairs of light
quarks are produced (see Fig.~\ref{B-psi}).

The interaction in the final state due to the annihilation results in the amplitudes
in the form
\begin{eqnarray}\label{Bs-decay}
    \mathcal M_s &\sim |s\bar s\rangle+ \mathcal A'\left\{|u\bar u\rangle+
    |d\bar d\rangle+\sqrt{\lambda_s}|s\bar s\rangle\right\},\\
    \label{B-decays}
    \mathcal M_d &\sim |d\bar d\rangle+ \mathcal A'\left\{|u\bar u\rangle+
    |d\bar d\rangle+\sqrt{\lambda_s}|s\bar s\rangle\right\},
\end{eqnarray}
for $B_s^0$ and $B^0$, respectively. Therefore, taking into account corrections
caused by differences in values of relative momenta $k[\eta]$, $k[\eta^\prime]$
in channels with $\eta$ and $\eta^\prime$ in the rest frame systems for
neutral $B$ meson, we easily find\footnote{In the limit of $\mathcal A'\to 0$
we arrive to formulae in \cite{Datta:2001ir}.}
\begin{equation}\label{Bs-rate}
    R_s=\frac{\Gamma[B_s^0\to J/\psi\eta^\prime]}{\Gamma[B_s^0\to J/\psi\eta]}=
    q_s
    \left|\frac{\sqrt{2}\mathcal A'\tan\phi+1+\mathcal A'\sqrt{\lambda_s}}
    {\sqrt{2}\mathcal A'-(1+\mathcal A'\sqrt{\lambda_s})\tan\phi}\right|^2
\end{equation}
and
\begin{equation}\label{B-rate}
    R_d=\frac{\Gamma[B^0\to J/\psi\eta^\prime]}{\Gamma[B^0\to J/\psi\eta]}=
    q_d
    \left|\frac{\frac{1}{\sqrt{2}}(1+\mathcal A')\tan\phi+\mathcal A'\sqrt{\lambda_s}}
    {\frac{1}{\sqrt{2}}(1+\mathcal A')-\mathcal A'\sqrt{\lambda_s}\tan\phi}\right|^2,
\end{equation}
where we define the factors
$$
    q_s=\left(\frac{k[B_s^0\to\eta^\prime]}{k[B_s^0\to\eta]}\right)^3,\qquad
    q_d=\left(\frac{k[B^0\to\eta^\prime]}{k[B^0\to\eta]}\right)^3,
$$
which are caused by differences in phase spaces and matrix elements squared,
which are proportional to relative momenta squared, since the decay takes
place in p-wave. Numerically, these factors give the correction equal to
20\%. In expressions (\ref{Bs-rate})--(\ref{B-rate}) the contribution of
annihilation amplitude $\mathcal A'$ is the parameter of model, because it
cannot be extracted, for instance, from the charge exchange reactions,
wherein there are the $t$-channel exchanges, which are absent in decays under
consideration.

\begin{figure}[ht]
\setlength{\unitlength}{.95mm}
\begin{center}
\begin{picture}(100,55)
\put(0,0){  \includegraphics[width=75\unitlength]{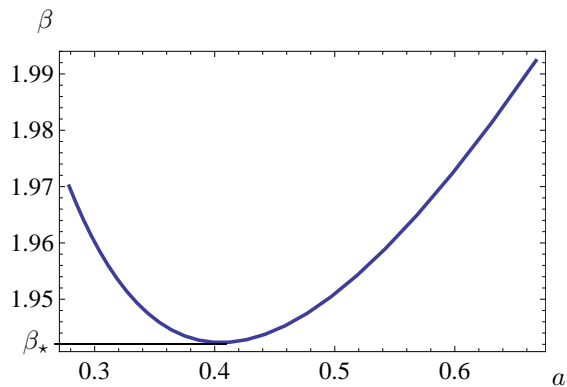}}
\put(5,50){$\beta$}
\put(77,0){$a$}
\put(3,5.){$\beta_\star$}
\put(7.5,5.8){\line(2,0){24}}
\end{picture}
\end{center}
  \caption{The parametric curve for the phase of $\beta$ versus the absolute
  value of relative amplitude for the annihilation of light quarks
  in the final state at the fixed value of $R_s$ (see (\ref{Bs-rate})--(\ref{B-rate})).
}
  \label{B_phase}
\end{figure}

\begin{figure}[t]
\setlength{\unitlength}{0.95mm}
\begin{center}
\begin{picture}(100,50)
\put(0,0){  \includegraphics[width=75\unitlength]{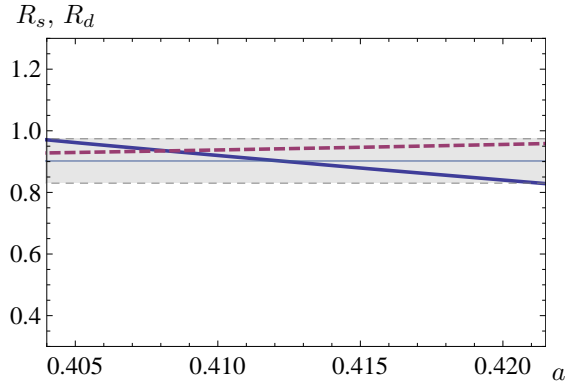}}
\put(2,50){$R_s,\,R_d$}
\put(77,0){$a$}
\end{picture}
\end{center}
  \caption{The ratio of $\eta^{(\prime)}$ yields in decays of neutral $B$
  mesons into $J/\psi$ versus the real parameter $a$ (see
  (\ref{Bs-rate})--(\ref{a-prime})). The solid line gives $R_s$, the dotted
  line presents $R_d$, the shaded band shows the experimental value of $R_s$
  with statistical uncertainties.
}
  \label{B-rates}
\end{figure}

The experimental data on $R_s$ \cite{Chang:2012gnb,Aaij:2014jna} have the
precision significantly better than that of $R_d$:  $R_s\approx 0.90\pm
0.09$, while $R_d\approx 1.11\pm 0.49$ (for the sake of simplicity of
consideration, here we have taken the averaged values for statistical and
systematic uncertainties of measurements). Under this fact we consider the
model with fixed values of $\tan\phi=0.75$ and $\lambda_s=0.5$ in order to
find a solution for  complex number $\mathcal A'=a\,\mathrm e^{i\beta}$ with
real parameters of $a$ and $\beta$ in the equation with the fixed value of
$R_s$ equal to its central value. The solution is presented in
Fig.~\ref{B_phase}, wherein one can see that the phase has the region of
stability at $\beta=\beta_\star$. In order to enlarge a predictive power of
model we have to aim a situation with a minimal possible spread of parameters
in the model. Fortunately, this is reached in the case of extremal point in
the region of correlations between the parameters: the phase can be fixed,
while the magnitude gets a minimal variation. Taking the value of phase in
the point of stability on Fig.~\ref{B_phase}, we show the prediction of model
for the ratio of yields in Fig.~\ref{B-rates}.

%%% Stable phase

Thus, the data on $R_s$ are well described within the statistical uncertainties if we put
\begin{equation}\label{a-prime}
    \mathcal A'=a\,\mathrm e^{i\beta_\star},\qquad a\in[0.4040;0.4215],
\end{equation}
i.e at the constant phase and changing real amplitude\footnote{At $\mathcal
A'\equiv 0$, i.e. in the case of switching off the mechanism of annihilation
in the final state, $R_s$ exceeds the experimental value by 70\%.} $a$.
Therefore, the model predicts
\begin{equation}{\label{revis9}}%$$
    R_d=0.943\pm0.015,
\end{equation}%$$
which is in a good
agreement with the experimental result, of course. The accuracy of prediction
is, at least, one order of magnitude less than the uncertainty of current
data.

\section{Conclusion}
We have studied the mechanism generating the $t$-dependence of ratio for the
cross sections of  $\eta'$ and $\eta$ mesons in reactions of charge exchange
for pions and kaons off protons with mixing of isoscalar components in
$\eta^{(\prime)}$. We have shown that such the dependence appears under the
introduction of interaction in the final state including the contribution of
annihilation.

The annihilation channel is the only one that contributes to the yields of
$\eta^{(\prime)}$ in radiative decays of $J/\psi$. This fact allows us to
estimate the mixing angle in consistency with the data on the charge exchange
reactions, if we take into account the uncertainty in the phenomenological value of
factor for the suppression of strange quarks. Then,
\begin{equation}\label{fin}
    \phi\approx 36.5\pm 0.8^o,
\end{equation}
that %significantly
improves the accuracy in the value of mixing angle and is
in agreement with results of other authors. The improvement of accuracy for
the parameter of suppression of strange quarks, breaking down the symmetry of
light quark flavors in the strong interactions, would allow us essentially to
reduce the uncertainty in the mixing angle of isoscalar quark states for the
case of neutral pseudoscalar mesons.

The account for the annihilation channel in the final state interaction
allows us to describe the ratio of $\eta^{(\prime)}$ yields in decays of
neutral $B$ mesons in the transition $B_{(s)}^0\to J/\psi$. In this
description the value of mixing angle is consistent with the constraints
obtained in the analysis for the charge exchange reactions. Then, accepting
the data on $B_s^0$ we have predicted the ratio of yields in decays of $B^0$
with the uncertainty one order of magnitude less than the experimental
uncertainty at present.

\acknowledgments This work is supported by Russian Foundation for Basic
Research, grant \# 15-02-03244.

\bibliography{bibmix-1}

\end{document}